\documentclass[sigconf]{acmart}
\AtBeginDocument{%
  }

\DeclareMathAlphabet{\pazocal}{OMS}{zplm}{m}{n}
\newcommand{\Lb}{\pazocal{L}}
\newcommand{\Tb}{\pazocal{T}}

\newcommand{\db}{\pazocal{D}}
\newcommand{\myparagraph}[1]{\vspace{3.0pt}\noindent{\textbf{\textit{#1:}}}~}

\usepackage{multirow}
\usepackage{algorithm}
 \usepackage{tikz}
  \usepackage{subfig}
\usepackage{algorithmic}
\setcopyright{acmlicensed}
\copyrightyear{2018}
\acmYear{2018}
\acmDOI{XXXXXXX.XXXXXXX}
\acmConference[SIGSPATIAL '25]{Make sure to enter the correct
  conference title from your rights confirmation email}{Nov 03--06,
  2025}{Minneapolis, MN}
\acmISBN{978-1-4503-XXXX-X/2018/06}

\usepackage{subfiles}

\begin{document}

\title{DeepMDV: Global Spatial Matching for Multi-depot Vehicle Routing Problems}

\author{Saeed Nasehi, Farhana Choudhury, Egemen Tanin, Majid Sarvi}
\email{{saeed.nasehibasharzad@student., farhana.choudhury@, etanin@, majid.sarvi@} unimelb.edu.au}
\affiliation{%
  \institution{University of Melbourne}
  \city{Melbourne}
  \state{Victoria}
  \country{Australia}
}

\renewcommand{\shortauthors}{Nasehi et al.}

\begin{abstract}
The rapid growth of online retail and e-commerce has made effective and efficient Vehicle Routing Problem (VRP) solutions essential. To meet rising demand, companies are adding more depots, which changes the VRP problem to a complex optimization task of Multi-Depot VRP (MDVRP) where the routing decisions of vehicles from multiple depots are highly interdependent. The complexities render traditional VRP methods suboptimal and non-scalable for the MDVRP. In this paper, we propose a novel approach to solve MDVRP addressing these interdependencies, hence achieving more effective results. The key idea is, the MDVRP can be broken down into two core spatial tasks: assigning customers to depots and optimizing the sequence of customer visits. We adopt task-decoupling approach and propose a two-stage framework that is scalable: (i) an interdependent partitioning module that embeds spatial and tour context directly into the representation space to \emph{globally} match customers to depots and assign them to tours; and (ii) an independent routing module that determines the optimal visit sequence within each tour. Extensive experiments on both synthetic and real-world datasets demonstrate that our method outperforms all baselines across varying problem sizes, including the adaptations of learning-based solutions for single-depot VRP. Its adaptability and performance make it a practical and readily deployable solution for real-world logistics challenges. 
\end{abstract}



\maketitle

\section{Introduction}
\label{sec:introduction}

The rapid growth of online retail and e-commerce has significantly increased delivery requests in urban areas, with thousands being processed every minute~\cite{wen2021package}. In 2022, Manhattan experienced over 2.4 million daily delivery requests~\cite{ManhatanDelivery}, equating to more than 3,000 deliveries every minute during a 12-hour workday. This highlights the need for solutions that are not only effective but also capable of producing results in a few seconds. Even algorithms that take minutes to compute fall short in handling such massive, real-time scheduling demands, as delays can quickly cascade, causing significant disruptions across the entire logistics network. Moreover, in large cities like New York City, for example, over 30,000 delivery trucks are active, each covering more than 12,000 miles annually~\cite{NYCDeliveryStatus}. This highlights the critical need for optimize delivery routes, as even a small reduction in driving distances can significantly cut travel costs and greenhouse gas emissions (e.g., a 1\% reduction could save over 3.6 million miles of travel each year).

The Multi-Depot Vehicle Routing Problem (MDVRP)~\cite{lim2005multi} is a spatially complex extension of the classic Vehicle Routing Problem (VRP), both of which are known to be NP-hard. Real-world instances of the MDVRP arise in large-scale retail operations, such as Target, where goods are dispatched from multiple distribution centers or stores (i.e., depots) to the end customers by delivery vehicles.

While the traditional VRP focuses on computing efficient routes from a single depot for delivery to customers (adhering to delivery vehicle capacity constraints), the MDVRP introduces a multi-origin spatial dimension: it requires simultaneously identifying the most suitable depot for each customer's demand and optimizing the routing plans for a fleet of delivery vehicles. This creates strong interdependencies between routes originating from different depots because changes in one vehicle’s route can influence others, since all customers need to be served and vehicles have capacity constraints. 

MDVRP solutions are broadly categorized into two approaches based on their methodology: (meta-)heuristic approaches and machine learning-based solutions. The \mbox{(meta-)heuristic} approaches rely on well-established methods, such as local search~\cite{alinaghian2018multi,wang2024adaptive}, genetic algorithms~\cite{surekha2011solution}, ant colony optimization~\cite{stodola2020hybrid}, large neighborhood search~\cite{pisinger2019large}, and tabu search~\cite{paul2021designing}. These methods rely heavily on iterative exploration of the solution space and typically require impractically extensive computation and memory to deliver competitive solutions for large urban-scale networks.

In contrast, deep learning-based solutions use the representation learning capabilities of neural networks to directly map problem instances to (near-)optimal solutions. Although many learning-based approaches address VRP, a notable gap exists for MDVRP. Existing few solutions for MDVRP~\cite{arishi2023multi,berto2024routefinder,li2024multi}  struggle to generalize when the number of customers or depots differs from the training settings. They also face scalability issues on large instances, as they can only provide high-quality results for small-size synthetic problems with a fixed number of depots matching the training configurations. This limits problem size and also makes assumptions such as all products being available at every depot. Moreover, existing learning-based solutions for VRP~\cite{kwon2020pomo, ye2024glop, zheng2024udc, gao2024towards} cannot be effectively extended to solve the MDVRP as they suffer from several limitations (see Section 2.2 for details) such as assuming that each customer must be assigned to the nearest depot. In contrast, our flexible method allows any customer to be served from alternative depots if doing so results in a better solution.

Solving MDVRP involves three interrelated tasks: (i) spatially assigning each customer to the most appropriate depot, (ii) assigning customers into feasible vehicle tours under capacity constraints, and (iii) determining the optimal intra-tour visiting sequence. The first two steps require coordination across all tours originating from each depot, while the third focuses on local route optimization. Poor decisions in depot or tour assignment can lead to unbalanced loads, inefficient routes, and increased operational costs.

To handle these challenges, we propose a novel task-decoupled deep learning architecture that separates spatial allocation (depot and tour assignment) from routing (tour sequencing). This modular design allows the first module to handle strategic spatial matching and load balancing, while the second focuses on optimizing the visit order within each tour. \emph{This separation allows for good solution quality, scalability, and generalization characteristics,} making the approach more practical for real-world MDVRP scenarios.

In this paper, we present \textbf{DeepMDV}, the first learning-based framework with strong spatial generalization capabilities across large-scale MDVRP instances with varying numbers of depots which is highly applicable to real-world scenarios. DeepMDV employs a two-stage design consisting of an interdependent partitioner and an independent router. The partitioner dynamically assigns each customer to a depot and a tour autoregressively, matching one customer at a time to a tour originating from a depot. Once the tours are defined, the router independently optimizes the sequence of visits within each tour, producing a complete solution for MDVRP.

The partitioner employs a transformer with a decoder comprising two key layers: the Tour Selection and Local Context Generation Layer (TSLCGL) and the Node Selection Layer (NSL). TSLCGL globally evaluates all tours and selects one per cycle, while NSL assigns a customer to the selected tour. This approach faces two challenges: First, NSL's focus on matching customers to a single tour may overlook the states of others. Second, evaluating all customers and tours greatly increases memory usage and training time.

To address these challenges efficiently, we introduce local context generated by the TSLCGL for each customer, which, combined with the customer’s initial embedding, allows the NSL to refine its assessment of nearby customers. Those better suited for the current tour are prioritized, while those better aligned with other tours are deprioritized. \emph{This approach ensures that the state of all neighboring customers is assessed in the context of all available tours, rather than relying solely on the selected tour’s state (in contrast to existing solutions), leading to more globally informed decisions.}


Moreover, our method follows a three-step training process. First, the router model is trained and then, the partitioner is trained using the router to generate optimal sequences, which are used in the reward function. After partitioner training, the router is fine-tuned using the tours generated by the partitioner as input. \emph{This iterative procedure improves the router's ability to deliver high-quality solutions for cross-distribution customer locations within tours}.

Furthermore, we propose a novel formulation to estimate the optimal maximum number of tours prior to solving each instance. By integrating this estimate into the model, we constrain the search space in a principled way, \emph{which not only improves solution quality but also significantly accelerates model convergence during training.}

The contribution of this paper are threefold: i) DeepMDV is the first learning-based approach that generalizes to large-scale MDVRP. By combining an interdependent partitioner with an independent router and embedding tour information into the representation space, our method effectively solves large-scale MDVRP instances, even with a different number of depots than those used during training. Its versatility also makes it highly effective for single-depot VRP, establishing DeepMDV as a practical solution for real-world applications. ii) We propose a three-step training approach to enhance the routing model's performance by taking the tours generated by partitioner as its input for fine-tuning, which is novel for any VRP or MDVRP solution. 
iii) A comprehensive evaluation on both synthetic (uniform and skewed distributions) and real-world MDVRP instances, along with sensitivity analysis, ablation studies, and a case study with visualization on real data demonstrate the superiority, scalability, and closeness of DeepMDV to the optimal solution compared to existing approaches for both single-depot and multi-depot instances.

\section{Related work}

This literature review focuses on (meta-)heuristic and learning-based approaches designed for the MDVRP. Also, we discuss learning-based VRP solvers  and their adaptation to MDVRP in Section~\ref{VRPlearningreview}. While exact algorithms, such as branch-and-bound~\cite{contardo2014new}, integer linear programming~\cite{benavent2013multi}, and solvers like Gurobi~\cite{gurobi} can theoretically solve MDVRP, their extremely high computational demands~\cite{kim2024clustering} make them impractical for instances exceeding even 50 customers. 

\subsection{Heuristic and meta-heuristic methods}

Several (meta-)heuristic approaches~\cite{yu2011parallel,ombuki2009using,surekha2011solution,lahyani2019hybrid,escobar2014hybrid} have been proposed to solve the MDVRP. These methods follow a construction-destruction-improvement pattern, where an initial solution is iteratively refined until a time limit or iteration threshold is reached. Despite their effectiveness and interpretability, these methods often require substantial memory and computation times~\cite{hou2022generalize}, making them impractical for real-life and large-scale scenarios. 

\citet{yu2011parallel} developed a parallel algorithm based on Ant Colony Optimization (ACO) to enhance computational efficiency and solution quality for MDVRP. Berman et al.~\cite{ombuki2009using} applied Genetic Algorithms (GA) to address MDVRP, while Surekha et al.~\cite{surekha2011solution} further refined GA-based approaches by grouping customers based on their proximity to the nearest depot before optimizing the routes within each cluster using GA. \citet{vidal2022hybrid} combined local search with GA, leveraging the strengths of both techniques to produce high-quality solutions for VRP, which can also be applied to MDVRP. ~\citet{lahyani2019hybrid} introduced a hybrid adaptive large neighborhood search Integrated with improvement procedures to enhance solution quality. Tabu Search (TS) has also been used to solve MDVRP. \citet{cordeau1997tabu} proposed a TS with the random initial solution, while \citet{escobar2014hybrid} proposed an enhanced strategy and used a hybrid Granular TS that leverages various neighborhood and diversification strategies to improve the quality of the initial solutions.

\subsection{Learning-based methods}~\label{VRPlearningreview}

Although many learning-based solutions have been developed for VRP~\cite{kool2018attention, xin2021multi, kwon2020pomo, ye2024glop, gunarathna2022dynamic, hou2022generalize,zheng2024udc,gao2024towards,luo2023neural,mozhdehi2024efectiw}, there is a significant gap in learning-based methods addressing large-scale MDVRP instances. Arishi et al.~\cite{arishi2023multi} introduced a state-of-the-art Multi-agent Deep Reinforcement Learning (MADRL) transformer trained by policy gradients for solving the MDVRP. Building on this, \citet{li2024multi} proposed a multi-type attention mechanism tailored to the MDVRP. Very recently, RouteFinder~\cite{berto2024routefinder} presented a generalized framework capable of addressing various VRP variants, including MDVRP, and demonstrated better performance compared to \cite{li2024multi}. However, all these approaches exhibit limited generalization and struggle to effectively scale across different instance sizes or varying numbers of depots, making them less suitable for real-world deployment. For example, all these approaches reported their performance for only up to 100 customers, whereas our method scales effectively to 1000-customer instances and finds solutions within seconds.

While a straightforward strategy to solve MDVRPs is using a distance-based clustering method~\cite{surekha2011solution}, followed by a VRP solver~\cite{kwon2020pomo, ye2024glop, mozhdehi2024efectiw, hou2022generalize,gao2024towards,luo2023neural,zheng2024udc} - this will lead to suboptimal results. Such strategy neglects global optimality for matching customers to depots, leading to imbalanced workload distribution among depots and ineffective customer assignments near cluster boundaries. Hence, some vehicles may be underutilized, leading to increased total travel distance and lowering overall effectiveness. This highlights the necessity of designing a model specifically for MDVRP. We use these methods as baselines to benchmark DeepMDV's performance.

\section{Problem definition}

The VRP~\cite{toth2002vehicle} is an optimization problem aimed at finding efficient routes for a fleet of vehicles to meet customer demands while starting and ending at a depot and adhering to vehicle capacity limits. The MDVRP~\cite{lim2005multi} extends the traditional VRP by incorporating multiple depots, with vehicles departing from one of them and returning to the same one after completing their customer deliveries.

In this paper, we address the problem of serving a set of customers $U$ using a fleet of vehicles $V$ operating from multiple depots $D$. Each route in $R$ must begin and end at the same depot, with the goal of minimizing total travel distance. All customer demands must be met while adhering to vehicle capacity constraints. A complete formulation of the problem is provided in Appendix~\ref{app:problem_formulation}.

\subsection{Definition of tours}
Before presenting our solution, we first define key concepts used to model the MDVRP. Here, we define the concepts of standby, initiated, active, and inactive tours. A \textbf{standby tour} is a tour that has not yet been assigned any customers, representing a vehicle at the depot with full capacity. An \textbf{initiated tour} has at least one assigned customer and can accept more. Assigning a customer to a standby tour turns it into an initiated tour. \textbf{Active tours} is a list of standby or initiated tours. Although this list could include multiple tours per depot, allowing more than one active tour per depot reduces training efficiency without improving solution quality. Thus, we assume only one active tour per depot at a time, limiting the total to $|\db|$, each from a distinct depot. \textbf{Inactive tours} is a list of tours that no longer accept new customers, having selected the depot as their final stop. When an active tour becomes inactive, it is replaced by a standby tour from the same depot.

\subsection{Markov decision process model}\label{App:MDP}

We model our approach as a Markov Decision Process (MDP)~\cite{puterman1990markov}, a standard framework for modeling decision-making where outcomes are influenced by actions taken. An MDP is defined by a 4-tuple $M = \{S, A, T, R\}$, where $S$ is the state space, $A$ is the action space, $T$ represents the transition dynamics, and $R$ is the reward function. Each of these components is detailed below.

\myparagraph{State} State $s_t \in S$ consists of two components, $(H, \Phi^a_t)$. 
The first component, $H$, describes the embedding of nodes, expressed as $H = \{ h_0, ..., h_n\}.$ Each embedding includes a vector for the node's location and a scalar for customer demand, with depot demands set to zero. The second component, $\Phi^a_t$, represents the state of all active tours at step $t$, denoted as $\Phi^a_t = \{\phi^t_1, ..., \phi^t_m \}$ where $m \leq |\db|$. The state of each active tour at step $t$ is described by $\phi_i^t = \{h_{d_i}, h_{v_i^{(t-1)}}, c^t_i\}$ which includes the embedding of depot it started from ($d_i$), the embedding of the last node added to the tour ($v_i^{(t-1)}$), and remaining available capacity ($c^t_i$). 

\myparagraph{Initial state} The initial state is defined as $(H, \Phi_0)$, where $H$ represents the node embeddings, and $\Phi_0$ denotes the initial state of all active tours at step $t=0$. The state of each active tour at this step is described by $\phi_i^0 = \{h_{d_i}, h_{d_i}, C\}$, where $h_{d_i}$ is the embedding of the depot for the tour, and $C$ shows the maximum available capacity.

\myparagraph{Action} The action at step $t$ involves selecting a tour from active tours list and assigning a node to it. In other words, $a^t \in A$ is defined as $(\phi_i^t, n_j)$, meaning node $n_j$ is added to tour $\phi_i^t$.

\myparagraph{Transition} The transition rule is to update the state $s_t$ to $s_{t+1}$ based on performed action $a_t = (\phi_i^t, n_j)$. In this process, the last added node of the selected tour is changed to $n_j$, and the remaining capacity of the tour is updated as $c^{t+1}_i = c^t_i - d_{n_j}$. If the selected node is the depot, the current tour $\phi_i$ is moved to the list of inactive tours. If the total number of tours has not yet reached the optimal maximum (details in Section~\ref{sec:OptimalNumber}), a new standby tour originated from the same depot replaces it in the active tour list.

\myparagraph{Reward} The reward is computed after all customers are assigned to tours and is defined as the negative sum of the minimum Euclidean distances for all tours. Each tour starts and ends at a depot, so the reward reflects the total length of the minimum Hamiltonian cycle per tour.
Finding the shortest Hamiltonian cycle, i.e., solving the Traveling Salesman Problem (TSP), is NP-hard. However, by dividing the entire problem space into multiple tours, our method enables the use of heuristic or machine learning-based solutions specifically designed for TSP, resulting in efficient and effective method for calculating rewards.

\begin{figure}[t]
\begin{center}
\centerline{\includegraphics[width=\columnwidth]{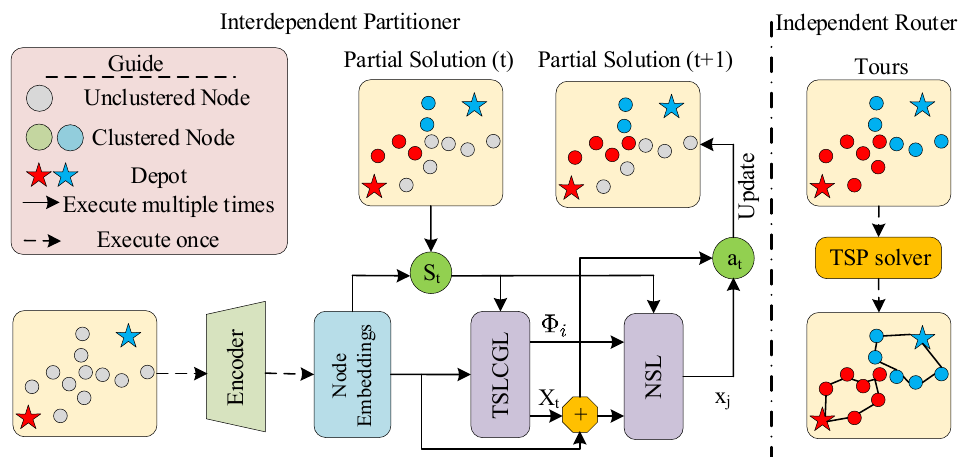}}
\vspace{-0.1in}
\caption{The DeepMDV uses the embeddings produced by encoder and the partial solution to generate local context and identify the candidate tour through TSLCGL. The node embedding, combined with the local context, used to choose the next best customer for the candidate tour. Once all customers are assigned, each tour is optimized using a TSP solver.}
\label{Fig:encoder_decoder_structure}
\end{center}
\vspace{-0.2in}
\end{figure}

\section{DeepMDV}
Figure~\ref{Fig:encoder_decoder_structure} shows the overview of the DeepMDV framework. The partitioner focuses on learning a stochastic policy $p_{\theta}(a_t|s_t)$, represented by a deep neural network with trainable parameter $\theta$. This policy partitions customers into multiple tours, ensuring each customer is assigned to one of tours by the end of the process. Each tour is associated with a specific depot, and the partitioner generates $m$ distinct groups of customers. The visiting sequence within each tour is then determined via routing policy $p'_{\theta}$ to minimize travel distance.

The partitioner policy $p_{\theta}$ is composed of an encoder and a decoder. The decoder includes a Tour Selection and Local Context Generation Layer (TSLCGL) along with a Node Selection Layer (NSL). As the problem instance remains fixed during the decision-making process, the encoder runs only once at the beginning, and its outputs are used in subsequent steps ($t>0$) for tour construction. 

At each step, the policy network selects a tour ($\phi^t_i$) via TSLCGL and a customer ($n_j$) for that tour through NSL, forming the action to update the partial solution and states. This continues until all customers are served. During training, we use AM~\cite{kool2018attention} as the TSP solver, while inference allows optimizing the visiting order within each tour using alternative methods like  LKH3~\cite{helsgaun2017extension}.

\subsection{Optimal maximum number of tours}\label{sec:OptimalNumber} In the MDVRP, each tour must adhere to vehicle capacity constraints to prevent exceeding the vehicle’s capacity. A new constraint can be defined to determine the optimal maximum number of tours before processing. Inspired by the approach proposed for the VRP~\cite{hou2022generalize}, we introduce this constraint for the MDVRP. For an MDVRP instance with $|U|$ customers and $|\db|$ depots, where $C$ represents the maximum vehicle capacity and $\delta_i$ denotes the demand of customer $i$, the optimal upper limit for the number of tours can be determined as $l_{max} = \lceil \frac{\sum_{i=0}^{|U|}\delta_i}{C} \rceil + |\db|$.

The first part of the equation determines the minimum number of tours needed to meet all demands. Since the optimal MDVRP solution may have up to $|\db|$ partially loaded tours, we add the number of depots to the minimum number of tours needed. This allows vehicles originating from each depot have the flexibility to deactivate an active tour before reaching their capacity limit, entails in a more efficient solution. $l_{max}$ represents the maximum number of tours allowed. Depending on the problem instance and the model, the actual number of tours can vary but will not exceed this limit. Our experiments show that predefining the optimal maximum number of tours accelerates model convergence. 

\myparagraph{Masking function for initiating a standby tour} To satisfy the optimal maximum number of tours constraint, the first step is to define a masking function that determines whether a standby tour can accept customers. A standby tour can turn into an initiated tour if the summation of the number of initiated tours and inactive tours is less than $l_{max}$.

\myparagraph{Masking function for deactivating an active tour} The second step to satisfy the optimal maximum number of tour constraint, is defining a masking function for deactivating an initiated tour. This function determines whether a tour $\phi_i$ with a total remaining capacity of $0< c_i \leq C$ can be deactivated by selecting the depot as the next stop. For this purpose, we define a threshold at decoding iteration $t$ named $\Tb_t$. This threshold will be used in Equation~\ref{eq:masked_prob} to determine whether a tour should be deactivated or not. Let $\phi_i$ denote an inactive tour with its unused capacity defined as wasted capacity $w_{\phi_i}$. Given the set of all inactive tours till last iteration ($\Lb_i^{(t-1)}$), the total wastable capacity at step $t$ and the threshold $\Tb$ at iteration $t$ is calculated as follow:
\begin{equation}
    \eta_t =l_{max} * C - \sum_{i=0}^{n}\delta_i - \sum_{\phi \in \Lb_i^{(t-1)}} w_{\phi}
\end{equation}
\begin{equation}\label{eq:threshold}
    \Tb_t =\frac{\eta_t}{l_{max} - |\Lb_i^{(t-1)}|}
\end{equation}

\subsection{Interdependent partitioning}
\subsubsection{Encoder} The encoder follows the architecture presented by~\citet{kool2018attention} and transforms customers and depots (call it nodes when we refer to both customers and depots) input $I_i$ into a hidden embedding $h_i$. The input of each customer $i$ comprises its coordinates $(x_i, y_i)$ and its associated demand $\delta_i$ while the value of demand for depots is zero. Unlike ~\citet{kool2018attention}, we transform node coordinates to polar coordinates relative to the first depot. Polar coordinates improve model generalizability by making representations invariant to spatial transformations like shifts, rotations, and scaling. This allows the model to capture relative positions and angular relationships between nodes, rather than relying on absolute locations. As a result, each node is represented by its relative Euclidean distance, polar angle, and demand.


\subsubsection{Decoder} Our proposed decoder model consists of two key layers: the Tour Selection and Local Context Generation Layer (TSLCGL) and the Node Selection Layer (NSL). At each iteration, the decoder is tasked with selecting a pair consisting of an active tour and a customer. The TSLCGL is responsible for identifying the tour with the highest `compatibility' (explained later) with the unvisited neighbor customers and generating local context. This context assists the NSL in choosing the next node for the selected tour, taking into account the status of other ongoing tours.

\begin{figure*}[t]
\begin{center}
\centerline{\includegraphics[width=\linewidth]{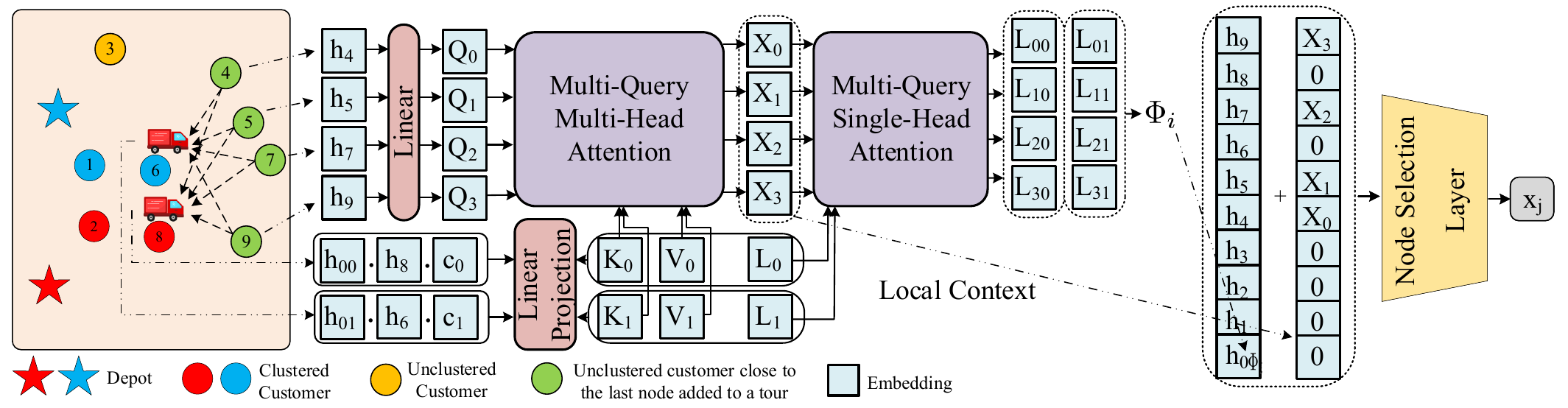}}
\caption{The architecture of decoder focusing on TSLCGL. First, the model selects the top $k=4$ nearest unvisited customers to the last nodes in active tours. Their embeddings pass through Multi-Query Multi-Head Attention, with keys and values derived from the active tour state, including the depot embedding, last node, and current capacity. This generates a local context for each customer, capturing the interaction with the active tours. This context is subsequently refined through a Multi-Query Single-Head Attention, which produces logits for selecting the best tour for the next node assignment. The local context, fused with initial embeddings, enables NSL to capture dynamic interactions between nodes and tours. Best viewed in color.}
\label{Fig:LCCGL_arch}
\vspace{-0.15in}
\end{center}
\end{figure*}

\myparagraph{TSLCGL} Figure~\ref{Fig:LCCGL_arch} illustrates the architecture of the proposed decoder, highlighting the TSLCGL. At each iteration, we identify the $k$ nearest nodes, denoted as $\zeta$, relative to the last customer added to each active tour. This approach aligns with findings that optimal actions in VRP are often concentrated among local neighbors~\cite{gao2024towards}. Then, the embeddings corresponding to these nodes in $\zeta$, are extracted from the encoder’s output and serve as queries in the multi-query, multi-head attention, where the keys and values come from each active tour's state:
\begin{equation}
    q_i = W^{Q_1}h_i, \: \: \forall \: i \in \zeta
\end{equation}
\begin{equation}
    k_j = W^{K_1}h_{\phi_j} \: , \: v_j = W^{V_1}h_{\phi_j} 
\end{equation}
where, 
\begin{equation}
   h_{\phi} = [h_{0\phi} , h_{l\phi} , c_{\phi}]
\end{equation}

Here $[$ · , · , · $]$ is the horizontal concatenation operator where $h_{0\phi}$ represents the embedding of the depot of the tour $\phi$, $h_l$ denotes the embedding of the last added node to the tour, and $c_\phi$ indicates the remaining capacity of the vehicle undertaking the tour. The local context is then calculated as:
\begin{equation}
    \pazocal{X}_i = softmax(\frac{q_ik^T_j}{\sqrt{dim_{k_j}}}) V_j
\end{equation}

The local context plays a vital role in guiding the NSL to prioritize nodes with a higher probability for a selected tour. It provides detailed information about the compatibility of each node with all active tours, \textbf{ensuring that the model avoids assigning nodes to a tour where they would be better suited for another one}. This strategic approach helps the model make more efficient decisions by optimizing the assignment of nodes to tours, thereby improving overall routing efficiency and \textbf{achieving towards a globally optimal solution}. 

In the final step, a multi-query single-head attention mechanism generates the unnormalized log-probabilities (logits) for each pair of tours and neighbor nodes. The queries and keys are defined as $q'_i = W^{Q_2}\pazocal{X}_i \: \:$, $\: \: v'_j = W^{K_2}h_{\phi_j}$. 

After using a max pooling, the logits for each tour are computed and clipped to the range $[-A, A]$ (with $A = 10$) using $tanh$ function. The tour with the highest probability is selected to pick the next node using $\phi = argmax(o)$.
\begin{equation}
    o = \begin{cases}
        A.tanh(Max(\frac{q'_i \: {v'_j}^T}{\sqrt{dim_{v'_j}}})), \hspace{0.2cm}   \: \text{if tour is active} \\
        -\infty, \hspace{2cm} \text{Otherwise}
    \end{cases}
\end{equation}

\myparagraph{NSL} Given the selected tour $\phi$ and local context $\pazocal{X}$ from TSLCGL, we apply attention mechanism to select the next node within the specified tour. The input embedding $h_i$ for this process consists of the tour depot's embedding and the updated node embeddings, which incorporate the local context. For non-depot nodes, each embedding is enriched by combining their local context with individual embeddings, effectively encoding the status of all active tours within each node's representation. 
\textbf{This strategy allows the model to consider the status of all active tours when selecting the next node, ensuring decisions are guided by overall tour dynamics}. So, the embedding $h_i$ is defined as follows:
\begin{equation}
    h_i = \begin{cases}
        h_{0\phi}, \: \: \: i = 0 \\
        h_i + \pazocal{X}_i, \: \: \: i \in \zeta_k \\
        h_i, \: \: \: Otherwise \\
    \end{cases}
\end{equation}

The query of the attention mechanism in NSL is defined as:
\begin{equation}
   q = W^{Q_3}[h_g , h_{0\phi} , h_{l\phi} , c_{\phi}]
\end{equation}
Where $h_g$ is the average value of the hidden embeddings of all nodes that are not previously visited, $h_{0\phi}$ is the depot node for this tour, $h_{l\phi}$ is the last added node to tour $\phi$, and $c_{\phi}$ shows the remaining capacity for tour $\phi$. Then, the key and value of the $i^{th}$ node is defined as $k_i=W^{K_3}h_{i}$, $v_i=W^{V_2}h_{i}$. 
Using attention mechanism, we compute the compatibility ($\mu_i$) between the query and all nodes.

To compute the output probabilities for visiting each node, we employ a final layer with a single attention head. In this layer, $\mu_i$ serves as the query and $h_i$ as the key. The compatibility scores are calculated and then clipped within the range [-A, A] (with A =10) using the $tanh$ function.

Now, we apply a masking function to enforce the threshold condition outlined in Equation~\ref{eq:threshold} while also handling visited nodes. The log-probabilities of selecting the depot as the next node, which would deactivate an initiated tour, is set to zero if the tour's current used capacity is below a defined threshold $\Tb_t$ by Equation~\ref{eq:threshold} for iteration $t$. If the tour's current capacity exceeds this threshold, the selection of the depot becomes conditional on the model's preference. So the compatibility score ($\mu_i^f$) and the probabilities ($p_i$) is calculated as follow:
\begin{equation}
    \mu_i^f = 
    \begin{cases}
        \mu_0, \hspace{1.2cm} \text{if i = 0 \& $C_{\phi} > \Tb_t$ }, \\
        \mu_i, \hspace{1.15cm}\text{if i $\neq$ 0 \& i is not visited}, \\
        -\infty, \hspace{2cm} Otherwise
    \end{cases}
    \label{eq:masked_prob}
\end{equation}
\begin{equation}
    p_i = p_{\theta} (\pi_t = I_i | s_t) = softmax(\mu_i^f)
\end{equation}

\subsection{Independent routing}
The independent routing module determines the optimal visiting order for nodes within each tour generated by the partitioner. Since customer demands are irrelevant to this step, the routing problem reduces to solving a TSP for each tour. Any TSP solver can be applied to address this subproblem, but heuristic methods often result in significantly longer training times for the partitioner. To address this, we choose AM~\cite{kool2018attention} for its simplicity and lower memory usage, making it ideal for efficient training.

\subsection{Three-step training}
We adopt policy gradient with rollout baseline~\cite{kool2018attention} to train the model. We define loss function as $\Lb(\theta|s) = \mathbb{E}_{p_{\theta}(\pi|s)} [L(\pi)]$ and optimize it using REINFORCE algorithm as follow:
\begin{equation}
    \nabla \Lb(\theta | s) = \mathbb{E}_{p_{\theta}(\pi|s)}\: [(L(\pi) - b(s)) \nabla \: log \: p_\theta(\pi|s)]
\end{equation}

The policy network $\pi_{\theta}$ generates probability vectors at each decoding step. A baseline (BL) with the same architecture is used as a greedy rollout baseline. The value of $\Lb(\theta|s)$ is calculated as the sum of the negative optimal length. $p_{\theta}$ and $p^{BL}_{\theta}$ is set to $p^p_{\theta}$ and $p^{BL_p}_{\theta}$ during partitioner training, and to $p^{AM}_{\theta}$ and $p^{BL_{AM}}_{\theta}$ otherwise. 

To enable parallel computation during router training or tour length evaluation, tours are padded by repeating the depot node to match the longest tour.
While AM trained on uniform distributions performs well on small MDVRP instances, its performance declines with scale-a limitation also noted by~\citet{hou2022generalize}-due to the deviation from uniform spatial distributions within each tour.

To tackle this, we employ a three-step training process: first, we train AM with a uniform distribution; once trained, we use it to generate reward values for the partitioner’s training. Finally, we apply the partitioner to solve a large-scale, randomly generated MDVRP dataset and fine-tune AM using the list of tours it generates. To manage the variability in the number of nodes per tour in each data batch, we implemented a padding technique that repeats the depot node to maintain a consistent number of nodes. This fine-tuning significantly improves AM's ability to generate \mbox{(near-)optimal} paths for large-scale MDVRP and VRP instances. The detailed training algorithm is presented in Appendix~\ref{app:training_alg}.

\section{Experimental evaluation}

\myparagraph{Hyperparameters and datasets} Following the setup of the previous works~\cite{kwon2020pomo, kool2018attention,hou2022generalize}, we sample the coordinates of $|\db|$ depots and $|U|$ customers uniformly from [0, 1] space, with customer demand randomly chosen between 1 and 10 units. Vehicle capacities are set to 50, 150, 175, and 200 units for instances with 100, 400, 700, and 1,000 customers, respectively. We generate 1,280,000 instances on the fly as training data. Training uses a batch size of 512 like previous approaches~\cite{kool2018attention,hou2022generalize}, reduced to 400 for instances with 200 nodes and 3 or 4 depots due to memory limits. We use a 6-layer encoder with 8 heads for multi-head attention and an embedding dimension of 128, with a constant learning rate $\eta = 10^{-4}$. 

We evaluate our method on three datasets: (i) the randomly generated dataset using a uniform distribution, used by default unless stated otherwise; (ii) a skewed dataset (see Sec. 5.2 for details); and (iii) a real-world dataset (see Sec. 5.3 for details).

\myparagraph{Baselines} We evaluate our method using two key metrics: objective value as total driving distance and runtime. Baselines include traditional solvers and state-of-the-art learning-based VRP and MDVRP methods: HGS~\cite{vidal2022hybrid}, OR-tools, GA~\cite{ombuki2009using}, 
POMO~\cite{kwon2020pomo}, GLOP~\cite{ye2024glop}, TAM~\cite{hou2022generalize}, LEHD~\cite{luo2023neural}, RouteFinder~\cite{berto2024routefinder}, UDC~\cite{zheng2024udc}, and MADRL~\cite{arishi2023multi}. Detailed description of the baselines are provided in Appendix~\ref{App:baselines}.

\myparagraph{Inference} During inference, the visiting sequence is obtained via LKH3 (CPU) or AM. We use DeepMDV trained on 100 nodes for 100-customer instances and on 200 nodes for larger ones. After a comprehensive sensitivity analysis (see appendix ~\ref{app:sensitivity} for details), we set the value of $k$ in TSLCGL 50 for 100-node instances and 30\% of the customer count for larger ones. We evaluate our method using four strategies: \textbf{i)~DeepMDV (AM, G)}: Utilizing DeepMDV with AM as the router with greedy search for both partitioning and routing. \textbf{ii)~DeepMDV (LKH3, G)}: Using the LKH3 for routing, while the partitioning is done using a greedy search approach. \textbf{iii)~DeepMDV (LKH3, G, P)}: employing multiple GPUs to run the greedy search in parallel with various values of $k$—specifically 30\%, 40\%, 50\% and 60\% of the number of customers. By leveraging the LKH3 for routing, we report the best result obtained for each instance. \textbf{iv)~DeepMDV (AM, S)}: applying DeepMDV with greedy AM for routing and sampling with a size of 1,000 for partitioning.

\subsection{Case study visualization}

\begin{figure}[t]
    \centering
        \vspace{-0.1in}
    \subfloat[Depots and customers location]{%
        \includegraphics[width=0.235\textwidth]{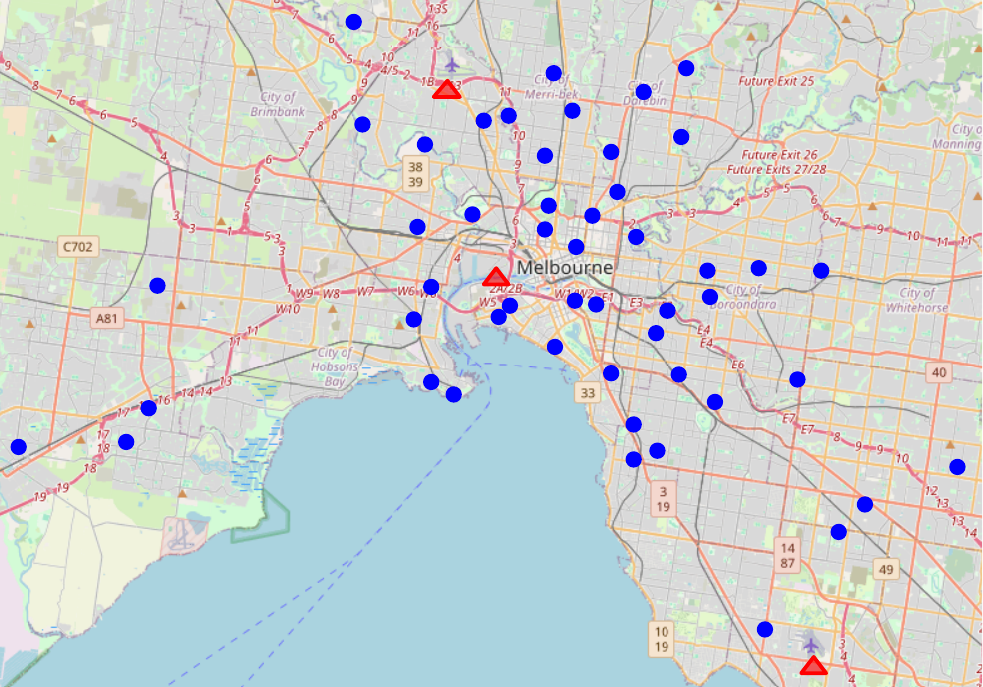}%
        \label{fig:sub1}} \hspace{0.1pt}
    \subfloat[Gurobi, Obj=5.1, RD = 326 km]{%
        \includegraphics[width=0.235\textwidth]{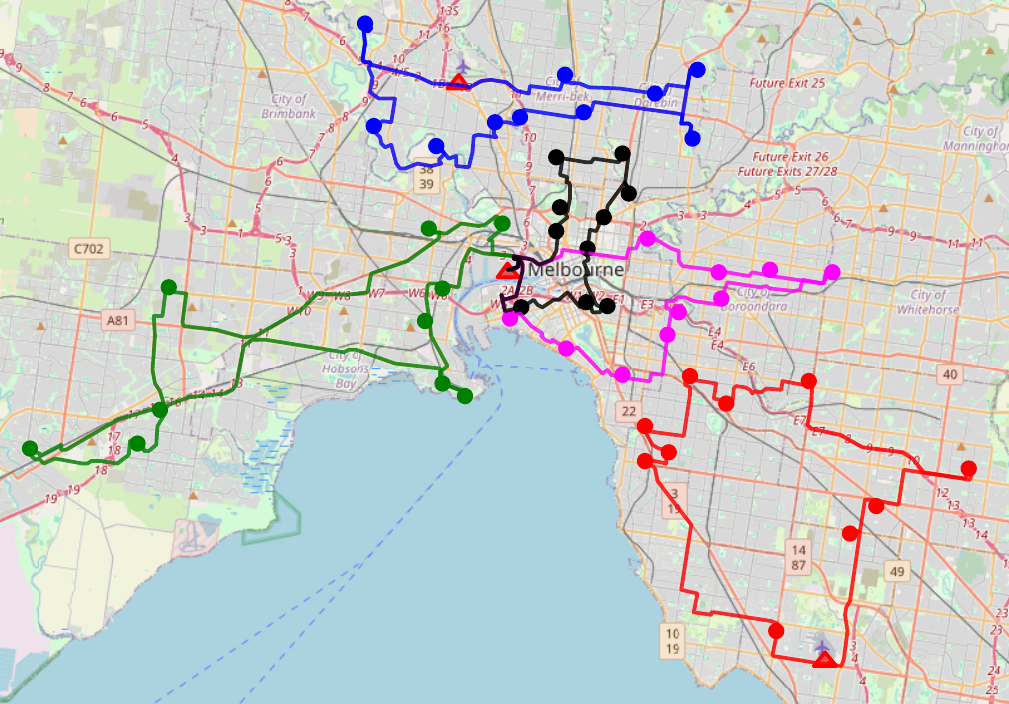}%
        \label{fig:sub2}}\\ \vspace{-0.1in}
    \subfloat[RouteFinder, Obj=5.4, RD = 351 km ]{%
        \includegraphics[width=0.235\textwidth]{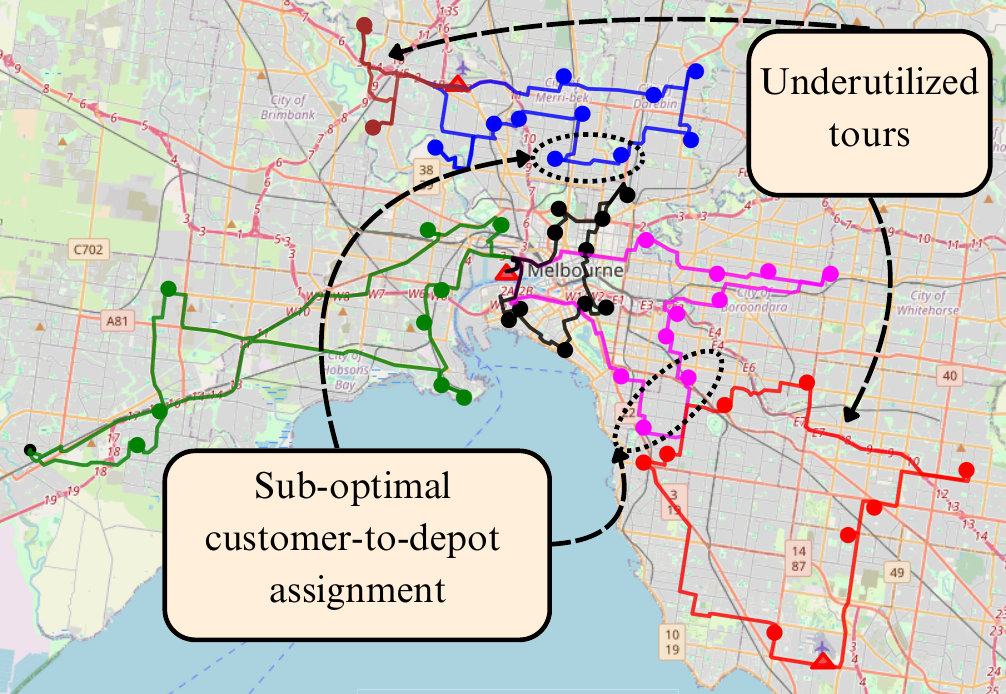}%
        \label{fig:sub4}}\hspace{0.1pt}
    \subfloat[DeepMDV, Obj=5.17, RD = 331 km ]{%
        \includegraphics[width=0.235\textwidth]{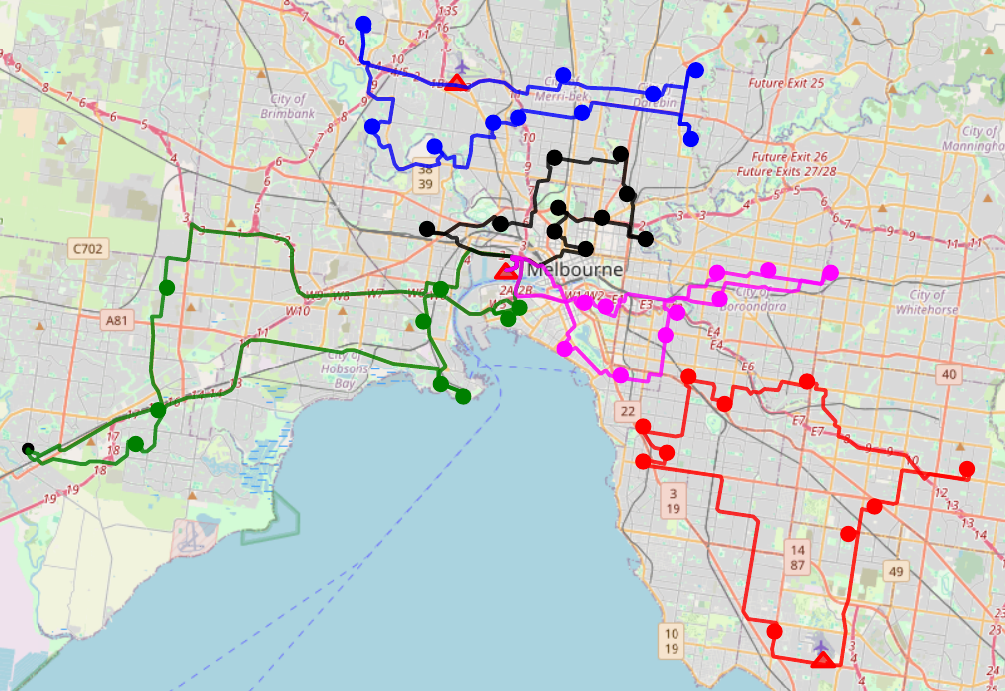}%
        \label{fig:sub3}}\vspace{-0.1in}
    \caption{MDVRP in Melbourne with three depots: two located near airports and one at a seaport, indicated by red triangles. Obj represents the Euclidean distance, while RD denotes the real driving distance on road network. While Gurobi and DeepMDV solve the problem using 5 tours, covering 326 km and 331 km respectively, Routefinder requires 6 tours and a total distance of 351 km to serve all customers.}
     \vspace{-0.2in}
    \label{fig:melbourne_MDVRP}
\end{figure}

We first present a case study visualization to highlight the importance of optimal depot assignment, using a real map of Melbourne with 50 customer locations and three depots (two airports, one seaport), shown in Figure~\ref{fig:melbourne_MDVRP}. Solutions are generated using three methods: Gurobi~\cite{gurobi}, which computes optimal solutions but takes several hours (as mentioned in Section 2); the proposed DeepMDV (AM,S) method; and RouteFinder, the latest baseline designed for the MDVRP. In terms of normalized objective (scaled to [0,1]), Gurobi achieves the lowest value at 5.10, followed closely by DeepMDV at 5.17, while RouteFinder trails with 5.4. When considering actual driving distances, Gurobi yields the shortest total route (326 km), followed by DeepMDV (331 km) and RouteFinder (351 km).

These results highlight the importance of an interdependent partitioning, which has been overlooked by prior methods. Its absence often leads to suboptimal customer-to-depot assignments, imbalanced tour workloads, and poor handling of customers near cluster boundaries (e.g., the four highlighted customers in Figure~\ref{fig:melbourne_MDVRP}c assigned to suboptimal depots). Hence, some vehicles are underutilized (red and brown tours in Figure~\ref{fig:melbourne_MDVRP}c), reducing overall efficiency.

Moreover, this case study demonstrates that even small improvements in normalized objectives translate to substantial real-world savings, even for small instances of only 50 customers. They also highlight DeepMDV’s effectiveness in accurately matching customers to depots and achieving near-optimal performance, significantly outperforming other learning-based methods like RouteFinder.

\begin{table*}[t]
\caption{Evaluation on a synthetic dataset. The objective (Obj.) is the driving distance, and the percentage gap (G) is relative to the HGS. Best value across methods is marked by ($^*$). For a fair comparison with traditional methods, all methods solve instances individually. Total runtime for 100 instances reported in the (T) column. The best results, excluding HGS, are bolded.}
\vspace{-0.12in}
\begin{center}
\begin{sc}
\begin{small}
\begin{tabular}
{p{3.5cm}|p{0.28cm}|p{0.6cm}p{0.6cm}p{0.6cm}|p{0.6cm}p{0.6cm}p{0.6cm}|p{0.6cm}p{0.7cm}p{0.6cm}|p{0.6cm}p{0.75cm}p{0.6cm}} 
\toprule
 \multirow{2}{*}{Methods} & & \multicolumn{3}{c|}{MDVRP100} & \multicolumn{3}{c|}{MDVRP400} &  \multicolumn{3}{c|}{MDVRP700} & \multicolumn{3}{c}{MDVRP1k}\\
 &$|\db|$& Obj. & G(\%) & T & Obj. & G(\%) & T & Obj. & G(\%) & T & Obj. & G(\%) & T \\
 \hline &&&&&&&&&&&&\\[-1em]
  HGS & \multirow{12}{*}{2} & 13.01$^*$ & 0.00 & 40m &  21.6$^*$ & 0.00 & 80m& 31.85 &0.00& 100m& 40.12 & 0.00 & 2h \\ 
  OR-tools & & 13.8 & 6.07 & 100m & 23.4 & 8.33 & 13h & 33.26 & 4.42 &20h& 40.77 & 1.62 & 27h \\
  GA & & 14.01 & 7.68 & 35m & 27.01 & 25.0 & 15h & - &- &- & - & -& - \\ 
  POMO & & 13.78 & 5.91 & 20s & 23.8 & 10.2 & 1m & 37.1&16.5 &2m& 54.6& 36.1 & 3m \\ 
  GLOP (LKH3) & & 15.8 & 21.4 & 40s & 24.3 & 12.5 & 2m & 33.2&4.23&3m&40.06& - 0.14 & 5m \\
  UDC & & 14.76& 13.45 & 30s & 24.02 & 11.2 & 2m &32.97 &3.51&3m&39.88& -0.5 & 4m \\
  RouteFinder & & 14.21&9.22 &30s & 24.4& 12.9& 2m&35.4 & 11.1&3m& 43.6& 8.7 & 4m\\
  MADRL & & 13.93 & 7.07 & 3m & 24.8 & 14.8 & 8m & 37.4&17.4 &12m& 50.3 & 25.4 &16m \\
  DeepMDV (AM, G) & & 13.84 & 6.37 & 45s & 22.76 & 5.37 &3m & 32.1 & 0.78 & 5m & 39.76 & -0.9 & 7m \\
  DeepMDV  (LKH3, G) & & 13.78 & 5.91 & 2m & 22.36 & 3.51 &4m & 31.56 & -0.91 & 6m & 39.03 & -2.71 & 8m \\
  DeepMDV (LKH3, G, P) & & 13.59 & 4.45 & 2m & \textbf{22.07} & \textbf{2.17} & 4m& \textbf{31.2}$^*$ & \textbf{-2.04} & 6m & \textbf{38.67}$^*$ & \textbf{-3.61} & 8m \\
  DeepMDV (AM, S) & & \textbf{13.34} & \textbf{2.53} & 2m & 22.25 & 3.00 &14m & - & - & - & - & - & - \\ 
  \hline &&&&&&&&&&&&&\\[-1em]
  HGS & \multirow{12}{*}{3} & 11.87$^*$ & 0.00 & 40m & 20.74$^*$ & 0.00 & 80m & 29.79&0.00&100m& 37.51 & 0.00& 2h \\ 
  OR-tools & & 12.66 & 6.65 & 100m & 22.05 & 6.31& 13h& 31.15&4.56&20h& 38.19 & 1.81 & 27h \\
  GA & & 12.88 & 8.50 & 30m & 25.88 & 24.8 & 14h & - & - & - & - & - & - \\ 
  POMO & & 12.91 & 8.76 & 20s & 21.96 & 5.88 & 1m & 33.0&10.8 &2m& 45.0& 19.9 & 3m \\ 
  GLOP (LKH3) & & 15.02 & 26.5 & 40s & 23.03 & 11.0 & 2m & 31.1&4.39&3m& 37.58& 0.18& 5m \\
  UDC & &13.69 & 15.3 & 30s & 22.74 & 9.64 & 2m & 30.9&3.72&3m&37.42& -0.23 & 4m \\
  RouteFinder & &13.29 & 11.9& 30s& 23.34&12.53 & 2m&33.08& 11.04&3m &40.9&9.03 & 4m \\
  MADRL & & 12.96 & 9.18 & 3m & 23.84 & 14.9 & 8m & 35.9&20.5&13m & 47.5 & 26.6 &16m \\
  DeepMDV (AM, G) & & 12.82 & 8.00 & 45s & 21.58 & 4.05 &3m & 30.23 & 1.47 & 5m & 37.44 & -0.18 & 7m \\
  DeepMDV (LKH3, G) & & 12.76 & 7.49 & 2m & 21.18 & 2.12 &4m & 29.64 & -0.5 & 7m & 36.64 & -2.31 & 8m \\
  DeepMDV (LKH3, G, P) & & 12.57 & 5.89 & 2m & \textbf{20.94} & \textbf{0.96} &4m & \textbf{29.35}$^*$ & \textbf{-1.47} & 6m & \textbf{36.29}$^*$ & \textbf{-3.25} & 8m \\
  DeepMDV (AM, S) & & \textbf{12.23} & \textbf{3.03} & 2m & 20.97 & 1.1 & 17m & - & - & - & - & - & - \\
  \hline &&&&&&&&&&&&&\\[-1em]
  HGS & \multirow{12}{*}{4} & 11.06$^*$ & 0.00 & 40m &  19.9$^*$ & 0.28 &80m & 28.45 &0.00&100m& 36.33 & 0.00& 2h \\ 
  OR-tools & & 11.8 & 6.69 & 100m & 20.97 & 5.37& 13h & 29.66& 4.25&20h& 36.21 & -0.33 & 27h \\
  GA & & 12.09 & 9.31 & 28m & 23.92 & 20.2& 13h & - & - & - & - & - & - \\ 
  POMO & & 12.07 & 9.13 & 20s & 21.15 & 6.28 & 1m & 30.44&6.99 &2m& 39.9& 9.82 & 3m \\ 
  GLOP (LKH3) & & 14.44 & 30.6 & 45s & 22.38 & 12.4 & 2m & 29.76&4.6 &3m& 36.64& 0.85& 5m \\
  UDC & &13.22 & 19.5 & 40s & 22.2 & 11.5 & 2m & 29.63&4.14&3m&36.51& 0.49 & 4m \\
  RouteFinder & & 12.15&9.85 & 30s& 22.46& 12.86& 2m& 31.22& 9.73&3m& 39.56& 8.89 & 4m\\
  MADRL & & 12.05 & 8.95 & 3m & 23.02& 15.7 & 8m &35.15&23.5&13m& 45.9 & 26.3 &16m \\
  DeepMDV (AM, G) & & 11.95 & 8.04 & 45s & 20.93 & 5.17 &3m & 29.43 & 3.44 & 5m & 36.48 & 0.41 & 7m \\
  DeepMDV (LKH3, G) & & 11.89 & 7.5 & 2m & 20.49 & 2.96 &4m & 28.77 & 1.12 & 6m & 35.64 & -1.89 & 8m \\
  DeepMDV (LKH3, G, P) & & 11.73 & 6.05 & 2m & \textbf{20.18} & \textbf{1.4} & 4m& \textbf{28.37}$^*$ & \textbf{-0.28} & 6m & \textbf{35.13}$^*$ & \textbf{-3.3} & 8m \\
  DeepMDV (AM, S) & & \textbf{11.47} & \textbf{3.7} & 3m & 20.25 & 1.75 &20m & - & - & - & - & - & - \\  
\bottomrule
\end{tabular}   
\end{small}
\label{table:syntetic_dataset_results}
\end{sc}
\end{center}
\vspace{-0.11in}
\end{table*}

\subsection{Cross-scale evaluation}

Table~\ref{table:syntetic_dataset_results} reports average performance over varying numbers of customers and depots, evaluated on a dataset generated using a uniform distribution. Accordingly, DeepMDV surpasses all learning-based baselines, including MADRL, GLOP, UDC, and RouteFinder. DeepMDV~(AM, S) achieves the best learning-based results on MDVRP100, with average optimality gaps of 2.53\%, 3.03\%, and 3.7\% for two, three, and four depots, respectively, relative to HGS. For larger cases, DeepMDV surpasses even HGS, showing at least a 3.25\% improvement on MDVRP1K. Note that T represents the total runtime for 100 instances and DeepMDV delivers fast and practical runtimes: for example, with T = 8M on MDVRP1K, the solution for each instance with 1,000 customers is obtained in about 4.8 seconds. DeepMDV~(LKH3, G, P) runs four parallel processes with different $k$ values, gaining up to 1.5\% improvement over fixed $k$, though it requires multiple GPUs. Sampling-based decoding outperforms greedy search on small instances but with higher runtime, and is omitted for 700+ customers due high computation time.

While HGS yields the best solutions for instances with up to 400 customers, it becomes computationally expensive and infeasible for larger problems. GA performs well on small instances but suffers from memory and runtime issues beyond 400 customers.

Among learning-based baseline methods, POMO performs well for smaller cases but quickly loses effectiveness as the problem size grows. Similarly, MADRL fails to generalize beyond 100 customers. UDC and GLOP perform better on large-scale problems but are less effective on small instances. RouteFinder scales better than previous baselines but still lags behind DeepMDV by at least 8\%, with the gap widening to over 13\% on larger instances.

\begin{table*}
\centering
\caption{Empirical results for instances with spatially skewed customer locations generated by Beta and Gamma distribution. The objective (Obj.) represents the driving distance, and the percentage gap (G) is relative to the HGS. The best value among all methods is marked with ($^*$), while the best results excluding HGS are highlighted in bold. Runtimes match those in Table~\ref{table:syntetic_dataset_results}, except for HGS, where they are extended to 40M, 100M, and 5H for instances with 100, 400, and 1,000 customers, respectively. }
\vspace{-0.12in}
\begin{center}
\begin{small}
\begin{sc}
\begin{tabular}{p{3.4cm}|p{0.3cm}|p{0.6cm}p{0.6cm}|p{0.6cm}p{0.6cm}|p{0.6cm}p{0.6cm}|p{0.6cm}p{0.6cm}|p{0.6cm}p{0.6cm}|p{0.6cm}p{0.6cm}} 
\toprule
 & & \multicolumn{6}{c|}{Beta Distribution} & \multicolumn{6}{c}{Gamma Distribution} \\[0.2em]
 \multirow{2}{*}{Methods} & & \multicolumn{2}{c|}{MDVRP100} & \multicolumn{2}{c|}{MDVRP400} & \multicolumn{2}{c|}{MDVRP1k}& \multicolumn{2}{c|}{MDVRP100} & \multicolumn{2}{c|}{MDVRP400}& \multicolumn{2}{c}{MDVRP1k}\\[0.2em]
 &$|\db|$& Obj. & G(\%) & Obj. & G(\%) & Obj. & G(\%) & Obj. & G(\%) & Obj. & G(\%)& Obj. & G(\%)\\
 \hline &&&&&&&&&&&&\\[-1em]
  HGS & \multirow{8}{*}{2} & 17.1$^*$& 0.00 & 25.8$^*$& 0.00 & 45.9$^*$ &0.00& 18.7$^*$ & 0.00 & 28.0$^*$ & 0.00& 49.6$^*$& 0.00 \\ 
  POMO & & 18.0& 5.26 & 26.5& 2.71 & 52.1 &13.5& 19.6 & 4.81 & 29.8 & 6.42&58.4 & 17.7 \\ 
  GLOP & & 19.2& 12.3 & 27.2& 5.42 & 48.6 &5.9& 21.5 & 15.0 & 30.6 & 9.28& 54.5 & 9.87\\
  UDC & &18.6 & 8.77 & 26.8 & 3.87 & 48.4 & 5.44 & 20.7&10.7&30.1&7.5& 54.2 & 9.27 \\
  RouteFinder & &18.4 &7.6 &28.3 & 9.6& 51.0&11.1 &20.1 &7.5 &31.8&13.5 & 57.3 & 15.5\\
  MADRL & & 18.2& 6.43 & 29.4& 14.0 & 53.8 & 17.2&19.8 &5.88& 32.5 & 16.1 & 59.8 & 20.6\\
  DeepMDV (AM, G) & &18.1&5.84&27.1& 5.03 & 48.7 & 6.1 & 19.5&4.27& 29.0 &3.57& 51.9 & 4.63 \\
  DeepMDV  (LKH3, G) & & 17.9& 4.67 & 26.5& 2.71 & 47.7 & 3.92 &19.3&3.2& 28.5 &1.78& 50.8 & 2.41 \\
  DeepMDV (LKH3, G, P) & & \textbf{17.7}& \textbf{3.5} & \textbf{26.3}& \textbf{1.93} & \textbf{47.2} & \textbf{2.83} & \textbf{19.2}&\textbf{2.67}& \textbf{28.3} &\textbf{1.07}& \textbf{50.5} & \textbf{1.81} \\
  \hline &&&&&&&&&&&\\[-1em]
  HGS & \multirow{8}{*}{3} & 15.2$^*$& 0.00 & 23.1$^*$ & 0.00 & 40.6$^*$&0.00& 17.8$^*$& 0.00 & 26.5$^*$ &0.00& 47.9$^*$ & 0.00 \\ 
  POMO & & 16.3& 7.23 & 24.7& 6.92 & 49.8 & 22.7 & 19.2&7.86& 28.4 & 7.16 & 56.5 & 17.8 \\ 
  GLOP & & 18.2& 19.7 &26.3& 13.9 & 45.9 & 13.1 & 20.4&14.6& 29.3 &10.6& 52.4 & 9.39 \\ 
  UDC & & 17.1& 12.5 & 25.8 & 11.7 & 45.4 & 11.8 & 19.5&9.55&28.9&9.05& 52.3 & 9.18 \\
  RouteFinder & &16.7 & 9.86&26.5 &14.7 & 48.3&19.0 &19.3 & 8.42&30.9& 16.6& 56.8 & 18.5\\
  MADRL & & 16.3& 7.23 &26.8& 16.0 & 51.1 & 25.9 & 19.2 &7.86& 31.3&18.1& 58.3 & 21.1 \\
  DeepMDV (AM, G) & & 16.2& 6.57 & 25.3& 9.52 & 45.4 & 11.8 &18.5&3.93& 27.8 &4.90& 50.5 & 5.42\\
  DeepMDV  (LKH3, G) & & 16.0& 5.26 & 24.8& 7.35 & 44.1 &8.62& 18.3& 2.80 & 27.4 &3.39& 49.3& 2.92 \\
  DeepMDV (LKH3, G, P) & & \textbf{15.8}& \textbf{3.94} & \textbf{24.5} & \textbf{6.06} &\textbf{43.2}& \textbf{6.40}&\textbf{18.2}& \textbf{2.24}& \textbf{27.2} &\textbf{2.64}&\textbf{49.0} & \textbf{2.29} \\
  \hline &&&&&&&&&&&\\[-1em]
  HGS & \multirow{8}{*}{4} & 15.1$^*$& 0.00 & 22.4$^*$& 0.00 & 39.8$^*$& 0.00&17.3$^*$& 0.00 &25.8$^*$ &0.00&  46.5$^*$ & 0.00 \\ 
  POMO & & 16.4& 8.61 &25.4& 13.4 & 47.3 & 18.8 &19.7& 13.9&28.6 &10.9& 55.7 & 19.8  \\ 
  GLOP & & 18.1& 19.9 &25.7& 14.7& 45.2 & 13.6 &20.1& 16.2& 29.1 &12.8& 52.1 & 12.0 \\
  UDC & & 17.4& 15.2 & 25.5 & 13.8 & 44.8 & 12.5 & 19.1&10.4&28.6&10.9& 51.7 & 11.2 \\
  RouteFinder & & 16.8&11.2 & 26.1& 16.5& 47.2& 18.5&18.9 &9.24 &29.8& 15.5& 54.4 & 17.0\\
  MADRL & & 16.7& 10.6 & 27.6& 23.2 & 53.2 &33.7& 18.7 & 8.09 & 30.6& 18.6& 56.4 & 21.2 \\
  DeepMDV (AM, G) & & 15.8& 4.63 & 24.3& 8.48 & 43.1 & 8.29 & 18.0& 4.04& 26.7 &3.48& 49.2 & 5.80\\
  DeepMDV  (LKH3, G) & & 15.7& 3.97 & 23.8& 6.25 & 42.0 & 5.52 &17.8&2.89& 26.4 &2.32& 48.2 & 3.65 \\
  DeepMDV (LKH3, G, P) & & \textbf{15.6}& \textbf{3.31} & \textbf{23.6}& \textbf{5.35}& \textbf{41.6} & \textbf{4.52} & \textbf{17.7}&\textbf{2.31}& \textbf{26.2} &\textbf{1.55}& \textbf{47.9} & \textbf{3.01} \\  \hline
\end{tabular}
\end{sc}
\end{small}
\end{center}
\vspace{-0.08in}
\label{table:skewedinstance}
\end{table*}

\subsection{Cross-distribution evaluation}
To further evaluate the method, we created a synthetic dataset with customers densely clustered and distant depots, simulating real-world logistics where hubs like airports or seaports serve as depots. Instances were constructed with fixed depot locations for: i) two depots: \{[0, 1], [1, 1]\}, ii) three depots: \{[0, 1], [0.5, 1], [1, 1]\}, iii) four depots: \{[0, 1], [0.33, 1], [0.66, 1], [1, 1]\}. Customer locations were generated using beta ($\mu=3$, $\sigma=1$) and gamma (shape $k=7$, scale $\theta=1$) distributions, normalized to the range [0, 1], to simulate urban settings where certain areas have high population density and demand, while others are more sparsely populated with fewer requests. For fair evaluation, all learning-based methods were fine-tuned on the dataset for one epoch. Following the sensitivity analysis (appendix ~\ref{app:sensitivity}), $k$ in TSLCGL was set to 60\% of the customers.

By skewing customer locations, we test the model’s ability to assign customers to the most suitable depot, even at a greater initial distance, to improve overall efficiency. Results are summarized in Table~\ref{table:skewedinstance}. HGS runtime was increased to 60 minutes for small and 6 hours for large instances to ensure competitive results.


As expected, previous methods performed poorly in this setting, favoring depot proximity and resulting in suboptimal routes with underutilized capacities. In contrast, DeepMDV achieved significantly better performance. Notably, while the gap between DeepMDV (LKH3, G) and UDC for normally distributed customer locations was less than 4\% in MDVRP1k, this gap widened to over 9\% in the skewed scenario. These findings highlight DeepMDV’s robustness and ability to optimize the full problem context across diverse spatial patterns in real-world, heterogeneous logistics settings.

\subsection{Results on a real-world dataset}\label{app:realworld} 

\begin{table*}[t]
\caption{DeepMDV vs. MADRL, the top baseline for small-scale problems, on real-world MDVRP dataset. The objective (Obj.) shows the driving distance, and the percentage gap (G) is w.r.t. the best known solution. Runtimes are reported in the column labeled (T). All methods solve each instance sequentially. For all instances, $k = 50\%$ of the number of customers.}
\vspace{-0.1in}
\begin{center}
\begin{sc}
\begin{small}
\begin{tabular}{p{1.3cm}|p{0.6cm}|p{0.4cm}|p{0.7cm}p{0.7cm}|p{1.4cm}p{1.4cm}p{1.4cm}p{1.4cm}p{1.4cm}p{1.4cm}}
\toprule
\multirow{2}{*}{Instance} & \multirow{2}{*}{|U|} & \multirow{2}{*}{$|\db|$}& \multicolumn{2}{c}{MADRL} & \multicolumn{2}{c}{\hspace{-1cm}DeepMDV (AM,G)} & \multicolumn{2}{c}{\hspace{-1cm}DeepMDV (LKH3, G)} & \multicolumn{2}{c}{\hspace{-1cm}DeepMDV (AM, S)} \\ &&&&&&&&&\\[-0.7em]
& & & G(\%) & T(s) & G(\%) & T(s) & G(\%) & T(s) & G(\%) & T(s)\\ \hline &&&&&&&&&&\\[-0.7em]
p04& 102 & 2 & 9.06 & $<2$ & 8.29 & $<1$ & 8.09 &   $<2$ & \textbf{4.96}&$<2$\\ [0.2em]
p05& 102 & 2 & 8.32 & $<2$ & 8.78 & $<1$ & 5.72 &   $<2$ & \textbf{4.59} & $<2$ \\ [0.2em]
p12& 82 & 2 & 10.34 & $<2$ & 10.39 & $<1$ & 10.08 & $<1$ & \textbf{6.75} & $<2$ \\ [0.2em]
p06& 103 & 3 & 9.32 & $<2$ & 8.14 & $<1$ & 7.33 &   $<2$ & \textbf{5.53} & $<2$ \\ [0.2em]
p01& 54 & 4 & 7.22 & $<1$ & 7.68 & $<1$ & 7.52 & $<1$ & \textbf{5.44} & $<2$\\ [0.2em]
p02& 54 & 4 & 10.67 & $<1$ & 8.69 & $<1$ & 8.52 & $<1$ &\textbf{4.89} & $<2$\\ [0.2em]
p07& 104 & 4 & 9.41 & $<2$ & 7.42 & $<1$ & 7.06 &   $<2$ & \textbf{4.82} & $<2$\\ [0.2em]
p15& 164 & 4 & 17.94 & $<2$ & 17.4 & $<1$ & 16.83 &  $<2$ & \textbf{10.81} & $<2$\\ \hline &&&&&&&&&&\\[-0.7em]
Average& - & - & 10.28 & $<2$ & 9.59 & $<1$ & 8.89 & $<2$ & \textbf{5.97}&$<2$\\ [0.2em] 
\bottomrule
\end{tabular}
\vspace{-0.07in}
\end{small}
\label{table:cvrplib}
\end{sc}
\end{center}
\end{table*}

To further assess DeepMDV’s effectiveness, we tested it against state-of-the-art solution on a real-world MDVRP dataset. This dataset includes multiple MDVRP instances, with scenarios involving up to 9 depots and hundreds of customers. We focused on instances with 2–4 depots, excluding those with added constraints like time windows. The selected instances are: p01, p02, p04, p05, p06, and p07 from Christofides et al.~\cite{christofides1969algorithm}, p12 and p15 from Chao et al.~\cite{chao1993new}. The results on this dataset are presented in Table~\ref{table:cvrplib}.

On average, MADRL exhibits a 10.28\% deviation from the Best Known Solution (BKS) across all instances. In contrast, DeepMDV (LKH3, G) delivers solutions in under 1 second, with an average gap of 8.89\%. DeepMDV (AM,S) reduces the gap to 5.97\% from the BKS, achieving results in under 2 seconds per instance, while the BKS reported in~\cite{sadati2021efficient} requires several hours of computation.

\begin{table*}[!th]
\caption{Performance of DeepMDV trained on two depots vs. baselines for VRP. The objective (Obj.) is driving distance, and gap (G) is relative to HGS. Best values across methods is marked by ($^*$). Run times are reported in the column labeled (T). All methods solve each problem instance sequentially, and we report the average runtime per instance. The best results, excluding HGS, are bolded. (†) denotes results taken from original papers due to unavailability of the source code.}
\vspace{-0.1in}
\begin{center}
\begin{sc}
\begin{small}
\begin{tabular}{p{3.5cm}|p{0.45cm}p{0.5cm}p{0.5cm}|p{0.45cm}p{0.7cm}p{0.6cm}|p{0.45cm}p{0.7cm}p{0.8cm}|p{0.45cm}p{0.7cm}p{0.8cm}} 
\toprule
 \multirow{2}{*}{Methods} & \multicolumn{3}{c|}{CVRP100} & \multicolumn{3}{c|}{CVRP400} &  \multicolumn{3}{c|}{CVRP700} & \multicolumn{3}{c}{CVRP1k}\\
 & Obj. & G(\%) & T & Obj. & G(\%) & T & Obj. & G(\%) & T & Obj. & G(\%) & T\\
 \hline &&&&&&&&&&&&\\[-1em]
  HGS & 15.5$^*$&0.00& 40M & 24.5$^*$ &0.00& 100M& 35.2$^*$ & 0.00 & 3h20M & 43.5$^*$ & 0.00 & 5H \\ 
  POMO & \textbf{15.7}&\textbf{1.29} &10s& 29.9 &22.0 &1m & 47.3 & 34.4& 2m & 101 & 232 & 3m\\ 
  TAM-AM † & 16.2 & 4.51 & 40s& 27.0&10.2 &3m& - & - & -& 50.1 & 15.2 & 5m\\
  TAM-LKH3 †  & 16.1 & 3.87 & 90s & 25.9 & 5.71&4m &-& - &- & 46.3 & 6.43 & 8m\\
  GLOP-LKH3 & 21.3 & 20.5 & 30s& 27.3&11.4&2m& 38.2& 8.52 & 3m & 45.9 & 5.51 & 3m\\
  LEHD & 16.2 & 4.5 & 40s& 25.6&4.48&3m& 36.8& 4.54 & 6m & 45.8 & 5.28 & 9m\\
  UDC & 17.4 & 12.25 & 30s& 26.2&6.93&2m& 37.4& 6.25 & 3m & 46.1 & 5.97 & 3m\\
  DeepMDV (AM, G) & 16.3 & 5.16 & 40s & 25.7 & 4.89 &3m & 36.8 & 4.54 & 4m & 45.5 & 4.59 &5m\\
  DeepMDV (LKH3, G) & 16.2 & 4.51 & 90s & 25.4 & 3.67 & 4m & 36.3 & 3.12 & 7m & 45.0 & 3.44 & 8m\\
  DeepMDV (LKH3, G, P) & 16.1 & 3.87 & 90s & \textbf{25.3} & \textbf{3.26} & 4m & \textbf{36.0} & \textbf{2.27}& 7m &\textbf{44.8} & \textbf{2.98} &8m\\
  \hline 
\end{tabular}
\vspace{-0.1in}
\end{small}
\end{sc}
\end{center}
\label{table:syntetic_dataset_results_single_depot}
\end{table*}


\begin{table*}[!hbt]
\caption{Performance of DeepMDV on instances with various depot numbers on a synthetic dataset. The objective (Obj.) represents the driving distance, and the percentage gap (G) is relative to the best value across all methods, marked by ($^*$). }
\vspace{-0.12in}
\begin{center}
\begin{sc}
\begin{small}
\begin{tabular}{p{2.78cm}|p{0.35cm}p{0.45cm}|p{0.35cm}p{0.58cm}|p{0.35cm}p{0.45cm}|p{0.35cm}p{0.58cm}|p{0.35cm}p{0.45cm}|p{0.35cm}p{0.53cm}|p{0.358cm}p{0.45cm}|p{0.35cm}p{0.53cm}} 
\toprule
 \multirow{3}{*}{Methods} & \multicolumn{2}{c|}{$|U|=100$} & \multicolumn{2}{c|}{$|U|=1k$}& \multicolumn{2}{c|}{$|U|=100$} & \multicolumn{2}{c|}{$|U|=1k$} & \multicolumn{2}{c|}{$|U|=100$} &  \multicolumn{2}{c|}{$|U|=1k$} & \multicolumn{2}{c|}{$|U|=100$}& \multicolumn{2}{c}{$|U|=1k$}\\
 & \multicolumn{2}{c|}{$|\db|$=1 } & \multicolumn{2}{c|}{$|\db|$=1, } & \multicolumn{2}{c|}{$|\db|$=2} & \multicolumn{2}{c|}{$|\db|$=2} & \multicolumn{2}{c|}{$|\db|$=3} & \multicolumn{2}{c|}{$|\db|$=3} & \multicolumn{2}{c|}{$|\db|$=4}& \multicolumn{2}{c}{$|\db|$=4}\\
 &Obj. & G(\%) & Obj. &G(\%)& Obj. & G(\%) & Obj. & G(\%) & Obj. & G(\%) & Obj. & G(\%) & Obj. & G(\%) & Obj. & G(\%) \\
 \hline &&&&&&&&&&&&&&\\[-1em]
  Trained on $|\db|=2$ & 16.2$^*$ & 0.00 & 45.0$^*$ &0.00 & 13.8$^*$ & 0.00 & 39.0$^*$&0.00 & 14.7& 14.8 & 39.2 & 7.1& 15.1 &26.9& 39.8& 11.8 \\ 
  Trained on $|\db|=3$ & 16.3 & 0.61 & 45.5 &1.11 & 13.8$^*$ & 0.00 & 39.5 &1.28 &12.8$^*$& 0.00 & 36.6$^*$ & 0.00 & 12.1 & 1.68 &36.0& 1.12  \\
  Trained on $|\db|=4$ & 16.5 & 1.85 & 46.2 &2.66 & 13.9 & 0.72 & 40.4&3.58 &12.8$^*$& 0.00 & 37.4 & 2.18 & 11.9$^*$ &0.00 &35.6$^*$& 0.00 \\ 
  \hline 
\end{tabular}
\end{small}
\end{sc}
\end{center}
\vspace{-0.12in}
\label{table:syntetic_MDVRP_Generalizability}
\end{table*}

\subsection{Cross-depot-size evaluation}
DeepMDV shows strong generalizability, not only across different customer scales but also varying depot counts, making it well-suited even for single-depot VRPs. Table~\ref{table:syntetic_dataset_results_single_depot} (for instances with fewer than 1,000 customers) and Table~\ref{table:syntetic_dataset_results_single_depot_further} in Appendix~\ref{app:large-scalevrp} (for instances exceeding 1,000 customers) present the results of DeepMDV, trained on MDVRP with two depots, compared to baselines specifically designed and trained for single-depot VRPs.

Using LKH3 as the router, our method outperforms all baselines on instances with over 400 customers and surpasses LEHD, GLOP, and UDC for CVRP1k by 1.7\%, 2\%, and 2.4\%, respectively . Running five parallel instances with varying $k$ values (e.g., 200, 300, 400, 500) further improves results, achieving gains of 2.22\%, 2.4\%, and 2.9\% over LEHD, GLOP, and UDC, respectively for CVRP1k.

Methods that require the number of depots in training and testing to match implicitly assume that all items are available at every depot, which is often unrealistic. DeepMDV’s ability to perform well under varying depot configurations makes it more practical for real-world applications where such assumptions may not hold.

Table~\ref{table:syntetic_MDVRP_Generalizability} further evaluates our model's adaptability to varying depot counts during testing. Although performance is best when training and testing depot numbers match, it remains strong when they differ. For instance, the model trained on three depots performs comparably to one trained on two depots for MDVRP100 with three depots, with similar trends for three and four depots.

DeepMDV trained on three depots generalizes well to different depot counts. For MDVRP100, it achieves gaps of 0.61\%, 0\%, and 1.68\% for instances with one, two, and four depots, respectively, compared to models trained on matching depot counts. Similarly, for MDVRP1000, the gaps are 1.11\%, 1.28\%, and 1.12\%. These small gaps highlight the model’s strong generalization capability across varying depot configurations without the need for retraining.

\subsection{Ablation study}\label{app:ablation}

\myparagraph{Three-step training} Our analysis shows that for MDVRP with two depots and 100 customers, the performance gap between the AM trained on a uniform distribution and LKH3 is under 0.5\%. However, as problem size increases to 400, 700, and 1000 customers, this gap grows to 9.5\%, 15.7\%, and 19.3\%, respectively. Training the AM with our proposed approach reduces these gaps to 1.34\%, 1.78\%, 1.71\%, and 1.87\%, respectively. This demonstrates that proposed approach significantly improves solution quality for larger instances, making it advantageous for models relying on AM as a secondary solver.

\myparagraph{Components in DeepMDV} We conduct studies to evaluate the impact of key components in our method: leveraging Local Context (LC), the Optimal Number of Tours (ONT), and Coordinate Transformation (CT). For each study, we train the model without one component while maintaining all other configurations. This allows us to quantify each component’s contribution to performance.  Results in Table~\ref{table:ablation} for MDVRP instances with 100 and 1000 customers and two depots validate the efficacy of each component's design.
\begin{table}[h]
\vspace{-0.05in}
\caption{Empirical results of ablation study. The gap \% is  w.r.t. the results with all components in use.}
\vspace{-0.14in}
\begin{center}
\begin{sc}
\begin{small}
\begin{tabular}
{p{1.8cm}p{1.8cm}|p{0.5cm}p{0.5cm}p{0.5cm}} 
\toprule
MDVRP100 & MDVRP1000 &  LC & ONT& CT \\ \hline &&&&\\
[-1em]
1.17\% & 2.9\% & $\times$ & & \\[0.2em]
0.57\% & 1.08\% & & $\times$ & \\[0.2em]
0.72\% & 2.61\% & &  & $\times$ \\[0.2em] \hline
\end{tabular}
\vspace{-0.1in}
\label{table:ablation}
\end{small}
\end{sc}
\end{center}
\end{table}

\myparagraph{Turn interdependent decision-making to independent}Instead of embedding tour information and using an interdependent partitioning approach, we could adopt a different strategy with separate models. For instance, \citet{li2021deep} propose a heterogeneous VRP method that first selects a vehicle and then assigns the next customer, without interdependent decision-making or considering other vehicles' states. Such an approach may lead to suboptimal solutions, as seen in Table~\ref{table:ablation} when local context is deactivated.

\subsection{Extended evaluations}
We conduct an extended evaluation to rigorously assess DeepMDV's memory consumption, performance, scalability, and generalization across a range of scenarios. Appendix~\ref{app:large-scalevrp} highlights DeepMDV's effectiveness on large-scale VRP instances, while Appendix~\ref{app:large-scale-MDvrp-Memory} compares its performance against HGS on large-scale MDVRP problems with 4 and 10 depots, focusing on objective quality, memory usage, and runtime efficiency. Across all experiments, DeepMDV delivers superior results, establishing itself as a robust and practical solution for both MDVRP and VRP in real-world scenarios.

We further conduct an extended evaluation to examine DeepMDV's sensitivity to the choice of the hyperparameter $k$ used in the algorithm. Setting $k$ too small relative to the number of customers results in suboptimal solutions across all tested distributions; however, performance stabilizes once a certain threshold is reached. Details are provided in Appendix~\ref{app:sensitivity}.

\section{Conclusion}

We present DeepMDV, a deep-learning method for solving large-scale Multi-Depot Vehicle Routing Problem (MDVRP). Leveraging attention mechanisms and a two-layer decoder, DeepMDV assigns customers to tours by first selecting the best tour and then the optimal customer within it. This task-decoupled structure enables effective spatial matching and sequencing, improving both accuracy and scalability. Experiments on both synthetic and real-world datasets demonstrate that DeepMDV consistently outperforms state-of-the-art learning-based MDVRP and VRP methods, efficiently solving large-scale instances with thousands of customers and multiple depots. It also generalizes effectively to varying numbers of depots, including those not seen during training, enabling for example strong performance on VRP without the need for retraining. Its ability to maintain solution quality under tight memory and time constraints highlights its practical viability. Furthermore, DeepMDV excels in skewed customer distributions where traditional methods struggle, making it well-suited for diverse urban logistics. 


\bibliographystyle{ACM-Reference-Format}
\bibliography{main}

\newpage
\appendix

\section{Problem formulation}\label{app:problem_formulation}
Here, we propose a mathematical formulation for MDVRP~\cite{surekha2011solution}. Let $V$ be a set of vehicles, where each vehicle $v \in V$ has a capacity $C_v$, and each customer $i\in U$ has a demand $\delta_i$. The distance between any two nodes $i$ and $j$, where $i,j \in \db \cup U $, is denoted by $e_{ij}$. We define $x_{ijv} \in \{0,1\}$ as a binary decision variable that equals 1 if vehicle $v$ travels from $i$ to $j$, and 0 otherwise. Let $z_j$ represent an auxiliary variable indicating the cumulative load of the vehicle after serving customer $j$, the MDVRP can then be formulated as follows:
\begin{equation}\label{eqn:1}
    Minimize \sum_{i \in \db \cup U} \sum_{j \in \db \cup U} \sum_{v\in V} e_{ij} x_{ijv}
\end{equation}
\begin{equation}\label{eqn:2}
    \sum_{v\in V} \sum_{i \in \db \cup U} x_{ijv} = 1, \: \forall j \in U
\end{equation} 
\begin{equation}\label{eqn:3}
    \sum_{i\in \db \cup U} x_{ijv} =  \sum_{j\in \db \cup U} x_{jiv}, \: \forall j \in U, \: \forall v \in V
\end{equation} 
\begin{equation}\label{eqn:4}
    \sum_{i \in U} x_{div} =1,  \sum_{i \in U} x_{idv} =1 , \: \forall d \in \db, \: \forall v \in V
\end{equation} 
\begin{equation}\label{eqn:5}
    \sum_{j\in U} \delta_j \sum_{i \in \db \cup U} x_{ijv} \leq C_v, \: v \in V
\end{equation}
\begin{equation}\label{eqn:6}
    z_j \geq z_i + n_i - M(i-x_{ijv}) , \: \forall i \neq j, \: \forall v \in V
\end{equation}

Equation~\ref{eqn:1} defines the objective of minimizing the total cost of all routes. Equation~\ref{eqn:2} ensures that each customer is visited exactly once, while Equation~\ref{eqn:3} enforces that any vehicle arriving at a customer must also depart from it. Equation~\ref{eqn:4} requires each vehicle’s route to start and end at a designated depot. Equation~\ref{eqn:5} limits each vehicle’s total served demand to its capacity, while Equation~\ref{eqn:6} eliminates subtours to ensure valid routes.

\section{Implementation details of baselines}\label{App:baselines}

\myparagraph{HGS~\cite{helsgaun2017extension}} We used the HGS implemented in PyVRP ~\cite{Wouda_Lan_Kool_PyVRP_2024} version 0.8.2 and configured the neighborhood size to 50, the minimum population to 50, and the generation to 100. The runtime was adjusted according to the size of the  MDVRP, choosing from 20, 30, 40, 50, or 60 seconds. All other parameters were left at their default.

\myparagraph{OR-tools} We use OR-tools with the PATH\_CHEAPEST\_ARC strategy as the initial solution and employ GUIDED\_LOCAL\_SEARCH for local search metaheuristics. The runtime was adjusted based on the size of the MDVRP instances, selecting from 60, 240, 480, 720, or 960 seconds to achieve an acceptable solution.

\myparagraph{Genetic Algorithm~\cite{ombuki2009using}} Following the original paper, the main parameters for the genetic algorithm were configured as follows: 500 generations, a crossover rate of 0.05, a mutation rate of 0.05, a route merge rate of 0.05, and a population size of 25.


\myparagraph{POMO~\cite{kwon2020pomo}} We first apply distance-based clustering~\cite{surekha2011solution}, followed by running the POMO with 8 augment inference for each cluster. 

\myparagraph{GLOP~\cite{ye2024glop}} Similar to the Cluster + POMO approach, this method utilizes GLOP with LKH3 as its TSP solver. We used the pre-trained model provided by the authors for evaluations. 

\myparagraph{UDC~\cite{zheng2024udc}} Similar to the Cluster + POMO approach, this method utilizes UDC to solve VRP at each cluster. We used the pre-trained model provided by the authors for evaluations. 

\myparagraph{LEHD~\cite{luo2023neural}} This baseline needs customer-size-specific inputs, making MDVRP integration impractical. We evaluate it only on single-depot VRP using the authors’ original code and pre-trained model.

\myparagraph{RouteFinder~\cite{berto2024routefinder}} This baseline finds solution for variety of VRP extentions including MDVRP. We leveraged the original implementation and used the pre-trained model provided by the authors. 

\myparagraph{MADRL~\cite{arishi2023multi}} We set the number of vehicles equal to the number of depots and use 2-opt after finding solution by greedy search. 

\section{Extended experiments on VRP}\label{app:large-scalevrp}

To further assess DeepMDV's scalability, we compare the performance of various methods on large-scale VRP instances with 2,000 and 7,000 customers. As shown in Table~\ref{table:syntetic_dataset_results_single_depot_further}, DeepMDV(LKH3, G) outperforms all baselines and it surpasses GLOP, the state-of-the-art VRP solver, for CVRP7k by approximately 1.54\%.Moreover, DeepMDV (LKH3, G, P) with $k$ values of 200, 300, 400, and 500 for CVRP2k and CVRP7k consistently outperforms all other methods by at least 2.45\%, 2.1\%, and 2.3\%, respectively.

\begin{table}[h!]
\caption{Performance of proposed method trained on MDVRP with two depots vs baselines for VRP on a synthetic dataset. The best value among all methods is marked with ($^*$).}
\vspace{-0.1in}
\begin{small}
\begin{center}
\begin{sc}
\begin{tabular}{p{2.8cm}|p{0.55cm}p{0.55cm}p{0.55cm}|p{0.55cm}p{0.6cm}p{0.55cm}}
\toprule
\multirow{2}{*}{Methods} & \multicolumn{3}{c|}{CVRP2k} & \multicolumn{3}{c}{CVRP7k}\\
 & Obj. & G(\%) & T & Obj. & G(\%) & T \\
 \hline &&&&\\[-1em]
  HGS &  62.9 &0.00& 7h& 212 & 0.00 & 10h \\
  AM & 114.3 & 81 &3m& 354 & 67.0 & 10m \\
  POMO & 485 &670 &8m & - & -& - \\ 
  TAM-AM & 74.3&18.1 &10m& 10.1 & 24.9 & 45m\\
  TAM-LKH3 & 64.8&3.02 &16m& 196.9 & -7.1 &1h \\
  GLOP-LKH3 & 63.0&0.1&3m&191.2& -9.81 & 10m \\
  UDC & 65.26&3.75&5m&188.2& -11.2 & 25m \\
  DeepMDV (AM, G) & 63.7 & 1.2 & 14m & 192.8 & -9.05 & 50m \\
  DeepMDV (LKH3, G) & 62.0 & -1.4 & 17m & 188.3 & -11.2 & 1h \\
  DeepMDV (LKH3, G, P) & \textbf{61.7$^*$} & \textbf{-1.9} & 17m & \textbf{186.9$^*$} & \textbf{-11.8} & 1h \\
  \hline 
\end{tabular}
\end{sc}
\end{center}
\vspace{-0.1in}
\end{small}
\label{table:syntetic_dataset_results_single_depot_further}
\end{table}

\section{Sensitivity analysis}\label{app:sensitivity}

\begin{figure*}[t]
\centering
  \includegraphics[width=\linewidth]{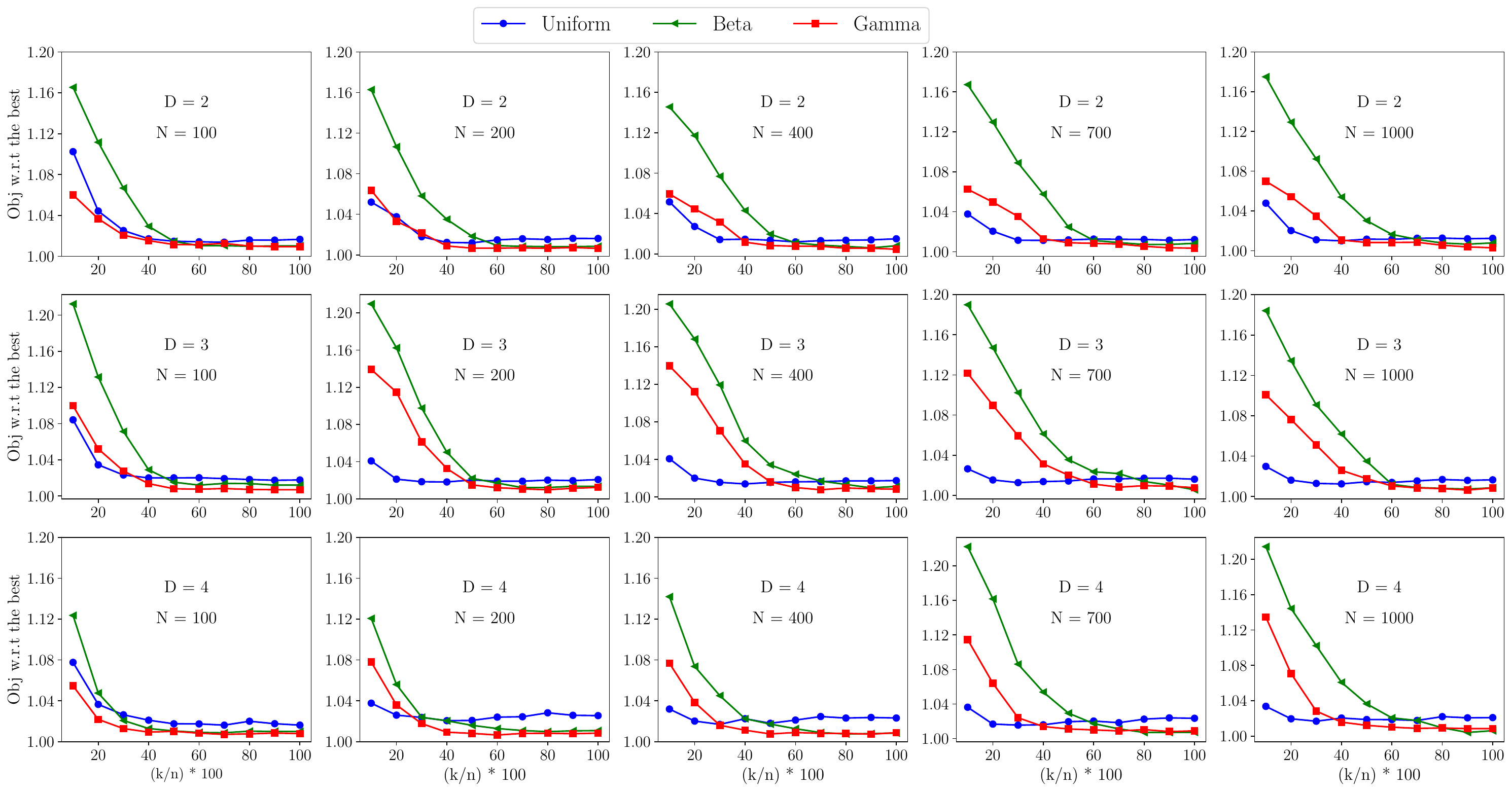}
\vspace{-0.3in}
  \caption{Sensitivity analysis of the influence of $k$ on solution quality, run on a synthetic dataset. The X-axis displays the value of $k$ as a percentage of the problem size, while the Y-axis shows the average cost relative to the best value achieved by the model which is calculated as the average of the minimum costs for each instance in the dataset, evaluated for different values of $k$. }
 \vspace{-0.15in}
\label{Fig:sensitivity}
\end{figure*}

To evaluate how the number of neighboring customers ($k$) affects solution quality, we tested DeepMDV with varying $k$ values. We define the best value for each problem instance as the minimum value obtained across all runs with different $k$ values. The average results across 100 problem instances, compared to the average of the best solutions, are presented in Figure~\ref{Fig:sensitivity}. 

Smaller $k$ values result in suboptimal results across all distributions, but performance stabilizes after a certain threshold; beyond that, larger $k$ only increases computation without improving quality. For large uniformly distributed instances, setting $k$ to 30\% yields strong results. However, setting $k$ to 50\% consistently produces high-quality solutions across distributions and customer sizes,, making it a promising choice for balancing efficiency and effectiveness.

\section{Scalability vs. Memory}\label{app:large-scale-MDvrp-Memory}

We designed an experiment to show that runtime isn’t the only bottleneck—traditional methods may still fail under memory constraints, even with ample computation time. We evaluated DeepMDV and HGS on 10 large-scale instances with 7,000 to 20,000 customers, 4 to 10 depots, and a vehicle capacity of 300. The results are presented in Table~\ref{table:syntetic_dataset_results_MDVRP_further}. DeepMDV, trained on 4-depot instances, was assessed based on runtime (T), total travel distance (obj), and memory usage (Mem). While HGS encounters an out-of-memory issue on a server with 110 GB of RAM, DeepMDV successfully solves the problem with 20,000 customers and 10 depots using less than 57 GB of memory, completing each instance in under 2 minutes. 

\begin{table}[h]
\caption{DeepMDV vs. HGS on large-scale MDVRP. }
\vspace{-0.1in}
\begin{small}
\begin{center}
\begin{sc}
\begin{tabular}{p{0.6cm}|p{0.25cm}|p{0.3cm}p{0.3cm}p{0.5cm}|p{0.4cm}p{0.4cm}p{0.4cm}|p{0.4cm}p{0.45cm}p{0.4cm}}
\toprule
\multirow{2}{*}{Scale} &\multirow{2}{*}{$\db$} & \multicolumn{3}{c|}{HGS} & \multicolumn{3}{c|}{DeepMDV(G,AM)} & \multicolumn{3}{c}{DeepMDV(G,LKH)}\\
 & & T. & Obj. & Mem & T. & Obj. & Mem & T. & Obj. & Mem \\
 \hline &&&&\\[-1em]
  7000 & 4&  2H &163& 33GB& 5M & 150 & 6GB & 6M & 144 & 6GB \\
  7000 & 10& 2H & 149 &37GB& 5M & 139 & 7GB & 6M & 132 & 7GB\\
  20000 & 4& 6H &- &OOM & 15M & 409& 39GB& 17M & 395 & 39GB\\ 
  20000 & 10& 6H&- &OOM& 17M & 394 &57GB & 19M & 376 & 57GB\\ \hline
\end{tabular}
\end{sc}
\end{center}
\vspace{-0.1in}
\end{small}
\label{table:syntetic_dataset_results_MDVRP_further}
\end{table}

\section{Detailed training algorithm}\label{app:training_alg}

Algorithm~\ref{alg:training} shows the three-step training procedure for DeepMDV. Lines 1–2 initialize inputs and iterate through three training steps. Lines 3–4 set the current and baseline policies depending on the step. Line 5 begins epoch-wise training. Lines 6–10 generate training data: TSP for step 1, MDVRP for step 2, and padded MDVRP solutions from the previous policy for step 3. Lines 12–13 sample solutions from the current and baseline policies. Lines 14–18 compute losses, with special handling and padding in step 2. Line 20 calculates the policy gradient, Line 21 updates the parameters, and Line 22 updates the baseline if improvement is observed.

\begin{algorithm}[H]
\footnotesize
   \caption{Three-step training algorithm with rollout baseline}
   \label{alg:training}
\begin{algorithmic}[1]
   \STATE {\bfseries Input:} Number of epochs $E_1 \& E_2 \& E_3$, batch size $B$
   \FOR {$step = 1$ {\bfseries to} $3$}
   \STATE \textbf{if} $step = 1$ \textbf{or} $3$ \textbf{then} $p_\theta^{BL} \gets p_\theta^{BL_{AM}}$, $p_\theta \gets p_\theta^{AM}$
   
   \textbf{else} $p_\theta^{BL} \gets p_\theta^{BL_{p}}$, $p_\theta \gets p_\theta^{p}$
   
   \FOR{$epoch=1$ {\bfseries to} $E$, $E${\bfseries selected from $E_1, E_2, E3$ based on the $step$}}
   \IF{$step = 1$}
   \STATE Generate Random TSP instances $s_i$, $\forall i \in \{ 1, ..., B\}$
   \ELSIF{$step =2$}
   \STATE Generate Random MDVRP instances $s_i$, $\forall i \in \{ 1, ..., B\}$
   \ELSE 
   \STATE Generate Random MDVRP instances, Solve by policy $p_\theta^p$ and pick routes $s_i$, $\forall i \in \{ 1, ..., B\}$, Padding all routes 
   \ENDIF
   \STATE Sample solution using policy $p_\theta$, $\forall i \in \{ 1, ..., B\}$ 
   \STATE Greedily generate solution using policy $p_\theta^{BL}$, $\forall i \in \{ 1, ..., B\}$
   
   \IF{$step = 1$ {\bfseries or} $3$}
   \STATE Calculate the loss $L(\pi_i)$, $L(\pi_i^{BL})$ $\forall i \in \{ 1, ..., B\}$
   \ELSE 
    \STATE Padding all routes of $\pi_i$ and $\pi_i^{BL}$, $\forall i \in \{ 1, ..., B\}$
   \STATE Calc the loss $L(\pi_i)$, $L(\pi_i^{BL})$ using $p_\theta^{AM}$, $\forall i \in \{ 1, ..., B\}$
   \ENDIF
   
   \STATE $\nabla \Lb \gets \frac{1}{B} \sum_{i=1}^B (L(\pi_i) - L(\pi_i^{BL})) \nabla_\theta \: log \: p_\theta(\pi_i)$
   
   \STATE $\theta \gets Adam(\theta, \nabla \Lb)$
  \STATE \textbf{if} {$p_\theta$ provides better result than $p_\theta^{BL}$} \textbf{then} $\theta^{BL} \gets \theta$
  \ENDFOR
\ENDFOR
\end{algorithmic}
\end{algorithm}

\section{Computational devices} We trained our model on an NVIDIA V100. Learning-based methods are tested on an NVIDIA V100 paired with an Intel Xeon Gold 6254 CPU (18 vCores), running Ubuntu 20.04 OS with 175 GB of Memory. Non-learning methods are executed on an AMD EPYC 9474F CPU with 28 vCores, using Ubuntu 20.04 and 110 GB of memory. 

\end{document}